\documentclass[12pt,manuscript]{emulateapj}

\def\simlt{\mathrel{\rlap{\lower 3pt\hbox{$\sim$}}
        \raise 2.0pt\hbox{$<$}}}
\def\simgt{\mathrel{\rlap{\lower 3pt\hbox{$\sim$}}
        \raise 2.0pt\hbox{$>$}}}
\newcommand{\target}{MAXI J1836--194}
\shorttitle{A variable compact jet in MAXI J1836--194}
\shortauthors{Russell et al.}

\begin{document}

\title{An evolving compact jet in the black hole X-ray binary MAXI J1836--194\altaffilmark{a}}

\author{
D. M. Russell\altaffilmark{1,2},
T. D. Russell\altaffilmark{3},
J. C. A. Miller-Jones\altaffilmark{3},
K. O'Brien\altaffilmark{4},
R. Soria\altaffilmark{3},
G. R. Sivakoff\altaffilmark{5},\\
T. Slaven-Blair\altaffilmark{3},
F. Lewis\altaffilmark{6,7},
S. Markoff\altaffilmark{8},
J. Homan\altaffilmark{9},
D. Altamirano\altaffilmark{8},
P. A. Curran\altaffilmark{3},
M. P. Rupen\altaffilmark{10},\\
T. M. Belloni\altaffilmark{11},
M. Cadolle Bel\altaffilmark{12},
P. Casella\altaffilmark{13},
S. Corbel\altaffilmark{14},
V. Dhawan\altaffilmark{10},
R. P. Fender\altaffilmark{15},
E. Gallo\altaffilmark{16},\\
P. Gandhi\altaffilmark{17,18},
S. Heinz\altaffilmark{19},
E. G.  K\"ording\altaffilmark{20},
H. A. Krimm\altaffilmark{21,22},
D. Maitra\altaffilmark{16},
S. Migliari\altaffilmark{23},
R. A. Remillard\altaffilmark{9},\\
C. L. Sarazin\altaffilmark{24},
T. Shahbaz\altaffilmark{1,2}
\&
V. Tudose\altaffilmark{25}
}
\altaffiltext{A}{Based on observations collected at the European Southern Observatory, Chile, under ESO Programme IDs 087.D-0914 and 089.D-0970.}
\altaffiltext{1}{Instituto de Astrof\'isica de Canarias (IAC), E-38200 La Laguna, Tenerife, Spain; russell@iac.es}
\altaffiltext{2}{Departamento de Astrof\'isica, Universidad de La Laguna (ULL), E-38206 La Laguna, Tenerife, Spain}
\altaffiltext{3}{International Centre for Radio Astronomy Research - Curtin University, GPO Box U1987, Perth, WA 6845, Australia}
\altaffiltext{4}{Department of Astrophysics, University of Oxford, Keble Road, Oxford OX1 3RH, UK}
\altaffiltext{5}{Department of Physics, University of Alberta, CCIS 4-181, Edmonton, AB T6G 2E1, Canada}
\altaffiltext{6}{Faulkes Telescope Project, University of Glamorgan, Pontypridd, CF37 1DL, UK}
\altaffiltext{7}{Department of Physics and Astronomy, The Open University, Walton Hall Milton Keynes, MK7 6AA, UK}
\altaffiltext{8}{Astronomical Institute `Anton Pannekoek', University of Amsterdam, PO Box 94249, 1090 GE Amsterdam, the Netherlands}
\altaffiltext{9}{MIT Kavli Institute for Astrophysics and Space Research, 70 Vassar Street, Cambridge, MA 02139, USA}
\altaffiltext{10}{NRAO Domenici Science Operations Center, 1003 Lopezville Road, Socorro, NM 87801, USA}
\altaffiltext{11}{INAF - Osservatorio Astronomico di Brera, Via E. Bianchi 46, I-23807 Merate (LC), Italy}
\altaffiltext{12}{European Space Agency, European Space Astronomy Centre, ISOC, Villa\~nueva de la Ca\~nada, Madrid, Spain}
\altaffiltext{13}{INAF - Osservatorio Astronomico di Roma, Via Frascati 33, I-00040 Monteporzio Catone (Roma), Italy}
\altaffiltext{14}{Laboratoire AIM, UMR 7158, CEA/DSM - CNRS - Universit\'e Paris Diderot, IRFU/SAp, Gif-sur-Yvette, France}
\altaffiltext{15}{School of Physics and Astronomy, University of Southampton, Southampton, Hampshire SO17 1BJ, UK}
\altaffiltext{16}{Department of Astronomy, University of Michigan, 500 Church Street, Ann Arbor, MI 48109, USA}
\altaffiltext{17}{ISAS, Japan Aerospace Exploration Agency, 3-1-1 Yoshinodai, chuo-ku, Sagamihara, Kanagawa 229-8510, Japan}
\altaffiltext{18}{Department of Physics, Durham University, South Road, Durham DH1 3LE, UK}
\altaffiltext{19}{Astronomy Department, University of Wisconsin-Madison, 475 N. Charter Street, Madison, WI 53706, USA}
\altaffiltext{20}{Department of Astrophysics/IMAPP, Radboud University Nijmegen, PO Box 9010, 6500 GL Nijmegen, the Netherlands}
\altaffiltext{21}{NASA/Goddard Space Flight Center, Greenbelt, MD 20771, USA}
\altaffiltext{22}{USRA, 10211 Wincopin Circle, Suite 500, Columbia, MD 21044, USA}
\altaffiltext{23}{Departament d'Astronomia i Meteorologia, Institut de Ci\`ences del Cosmos (ICC), Universitat de Barcelona (IEEC-UB), Mart\'i i Franqu\`es 1, E-08028 Barcelona, Spain}
\altaffiltext{24}{Department of Astronomy, University of Virginia, PO Box 400325, Charlottesville, VA 22904, USA}
\altaffiltext{25}{Institute for Space Sciences, Atomistilor 409, P.O. Box MG-23, Bucharest-M\v{a}gurele, RO-077125, Romania}

\begin{abstract}
We report striking changes in the broadband spectrum of the compact jet of the black hole transient \target\ over state transitions during its discovery outburst in 2011. A fading of the optical--infrared (IR) flux occurred as the source entered the hard--intermediate state, followed by a brightening as it returned to the hard state.
The optical--IR spectrum was consistent with a power law from optically thin synchrotron emission, except when the X-ray spectrum was softest.
By fitting the radio to optical spectra with a broken power law, we constrain the frequency and flux of the optically thick/thin break in the jet synchrotron spectrum. The break gradually shifted to higher frequencies as the source hardened at X-ray energies, from $\sim 10^{11}$ to $\sim 4 \times 10^{13}$ Hz. The radiative jet luminosity integrated over the spectrum appeared to be greatest when the source entered the hard state during the outburst decay (although this is dependent on the high energy cooling break, which is not seen directly), even though the radio flux was fading at the time. The physical process responsible for suppressing and reactivating the jet (neither of which are instantaneous but occur on timescales of weeks) is uncertain, but could arise from the varying inner accretion disk radius regulating the fraction of accreting matter that is channeled into the jet.
This provides an unprecedented insight into the connection between inflow and outflow, and has implications for the conditions required for jets to be produced, and hence their launching process.
\end{abstract}


\section{Introduction}

Relativistic jets are ubiquitous features of black hole X-ray binaries (BHXBs). When a BHXB is accreting in a low/hard X-ray state \citep[hereafter the hard state; for state definitions see][]{bell10}, synchrotron radiation from compact jets is commonly detected as a flat or inverted radio spectrum and an excess of infrared emission \citep{corbfe02,fend06}. In the canonical soft state, both the radio and infrared emission are quenched, by at least a factor of a few hundred at radio frequencies \citep[e.g.,][]{corbet00,homaet05,russet11b}, and jets may be prevented from being produced altogether in this state.

The conditions required for launching jets and the mechanical launching process are fundamental unknowns in the field of accretion. The nature of the quenching/recovery of jets over state transitions provides essential constraints for jet models. However to date, few studies have suggested how the jet might evolve over these transitions \citep{corbet13b}. During the hard-to-soft state transition, the IR/optical/UV emission from the jet appears to fade first at the start of the transition \citep[e.g.,][]{homaet05,russet10,yanyu12}, while the flat radio spectrum persists as the X-rays soften then finally fades before bright optically thin radio flares are observed, often associated with relativistic ejecta \citep*[e.g.,][]{fendet09,millet12}. Similarly, on the reverse transition on outburst decline, the IR emission from the jet appears to return after the radio emission \citep[e.g.,][]{coriet09,russet10,millet12,corbet13b}.

Phenomenologically, the detailed interplay between the well-documented pattern of changes in the inflow over the transition \citep{bell10} and the outflow, remains unclear. In the hard state the flat radio spectrum breaks to an optically thin power law around the infrared regime \citep[e.g.,][]{corbfe02,miglet10}. The break frequency varies between sources and even in time for the same source \citep{gandet11,russet13}. It is unknown if, or how the jet break varies during state transitions, but knowledge of this would provide essential constraints for models of how jets are produced.

\target\ was discovered in 2011 August as a new X-ray transient \citep{negoet11}. Within days it was classified as a BHXB due to its X-ray temporal/spectral properties and its bright jet seen at radio--IR frequencies \citep*{strosm11,millet11,russet11a,reiset12}.
The source underwent a state transition from the hard state to the hard--intermediate state \citep[HIMS;][]{bell10}, but instead of softening further it returned to the hard state and faded towards quiescence \citep{ferret12}. This type of `failed transition' outburst has now been reported in at least four BHXBs \citep[e.g.,][]{capiet09,ferret12,soleet13}.

Here we present the evolution of the jet spectrum of \target\ during its 2011 outburst using quasi-simultaneous multi-waveband radio, mm, mid-IR, near-IR, optical, UV and X-ray data.

\section{Broadband Observations}

\target\ was observed with the Karl G. Jansky Very Large Array (VLA) from 2011 September 3 to December 3.
We only consider data with quasi-simultaneous mm or mid/near-IR data available. The radio observations were split across frequencies from 1 to 43 GHz.
Due to the compact array configuration for most observations, we did not consider the lowest-frequency data ($<4$ GHz) as the target could not be easily resolved from a confusing source to the south west.
Data reduction was carried out according to standard procedures within the Common Astronomy Software Application (CASA).
External gain calibration was performed using the primary and secondary calibrators 3C286 and J1820-2528, respectively.  After phase-only self-calibration, the source flux densities were measured by fitting a point source in the image plane. Since no data were taken within three days of October 27, we estimated the radio spectrum on that date by fitting the smooth decay at each frequency with an exponential function, and interpolating for the appropriate MJD. 

Submillimeter Array (SMA) observations were taken on 2011 September 13 and 15, at frequencies of 256.5 and 266.8 GHz.  We used 3C454.3 as a bandpass calibrator, Neptune to set the flux scale, and the phase calibrators 1911-201 and 1924-292.  Data were reduced using standard procedures within CASA. \target\ was significantly detected on September 13 ($69.7 \pm 6.9$ mJy at 256.5 GHz; $66.4 \pm 6.3$ mJy at 266.8 GHz; the total on-source time was $\sim 3.5$ hours), but poor weather conditions on September 15 prevented us from placing any meaningful constraint on the source brightness.

Mid-IR observations of \target\ were made with the Very Large Telescope (VLT) on four dates during its 2011 outburst.
The VLT Imager and Spectrometer for mid Infrared (VISIR) instrument on UT3 (Melipal) was used in small-field imaging mode.
Observations were performed in \emph{PAH1} (8.2--9.0 $\mu m$), \emph{SIV} (10.3--10.7 $\mu m$) and \emph{J12.2} (11.7--12.2 $\mu m$) filters and, on some dates, $K$-band (2.0--2.3 $\mu m$).
Half of the approximately one-hour observing time on each date was on source, due to the chop-nod mode.
The data were reduced using the VISIR pipeline. Raw images were recombined and sensitivities were estimated based on standard star observations taken on the same night.
For $K$-band, flux calibration was achieved using two $K \sim 12.4$ magnitude stars from the Two Micron All Sky Survey \citep[2MASS;][]{skruet06}, both lying $\sim 6$\arcsec\ from \target.

Optical images of \target\ were acquired on six dates spanning September 5--30
using the 2-m Faulkes Telescopes (FTs) North and South, located at Haleakala on Maui, USA and Siding Spring, Australia, respectively, as part of an ongoing monitoring campaign of X-ray binaries \citep{lewiet08}. Images in Bessell $B$, $V$, $R$ and Sloan Digital Sky Survey (SDSS) $i^{\prime}$-bands
were reduced and calibrated using methods similar to those applied to other X-ray binaries observed by the FTs \citep[e.g.,][]{cadoet11,russet11b}.
Flux calibration in $V$ and $B$-bands was achieved using the UltraViolet/Optical Telescope (UVOT) calibration below. Landolt standard stars and a number of other calibrated fields in $R$ and $i^{\prime}$-bands were used to calibrate the field in these filters,
adopting the transformation to SDSS $i^{\prime}$-band from $R$ and $I$ described in \cite*{jordet06}. 

In the X-ray band, \textit{Swift} and the Rossi X-ray Timing Explorer (\textit{RXTE}) monitored \target\ every few days during its outburst \citep[see][]{ferret12}. UVOT on board \textit{Swift} observed in six optical/UV filters. {\tt uvotimsum} was used to combine images in the same filter 
and {\tt uvotsource} was used to perform aperture photometry.
Magnitudes are based on the UVOT photometric system, which differs from the Bessell system by V$-v < 0.04$ and B$-b < 0.04$ for all reasonable spectral indices \citep{poolet08}. We thus treat UVOT $v,b$ and FT $V,B$ magnitudes the same.
Following the procedure outlined by \cite{altaet08}, we calculated colors and intensities using the 16s time-resolution PCA Standard 2 mode data. For \target\ we defined hard and soft color as the ratio of count rates in the 16--20 and 6--16 keV bands, respectively, to that in the 2--6 keV band, and the intensity as that in 2--20 keV band. All values were averaged for each observation, and normalized to the Crab nebula on a per PCU basis.

\begin{figure}
\centering
\includegraphics[width=8.9cm,angle=0]{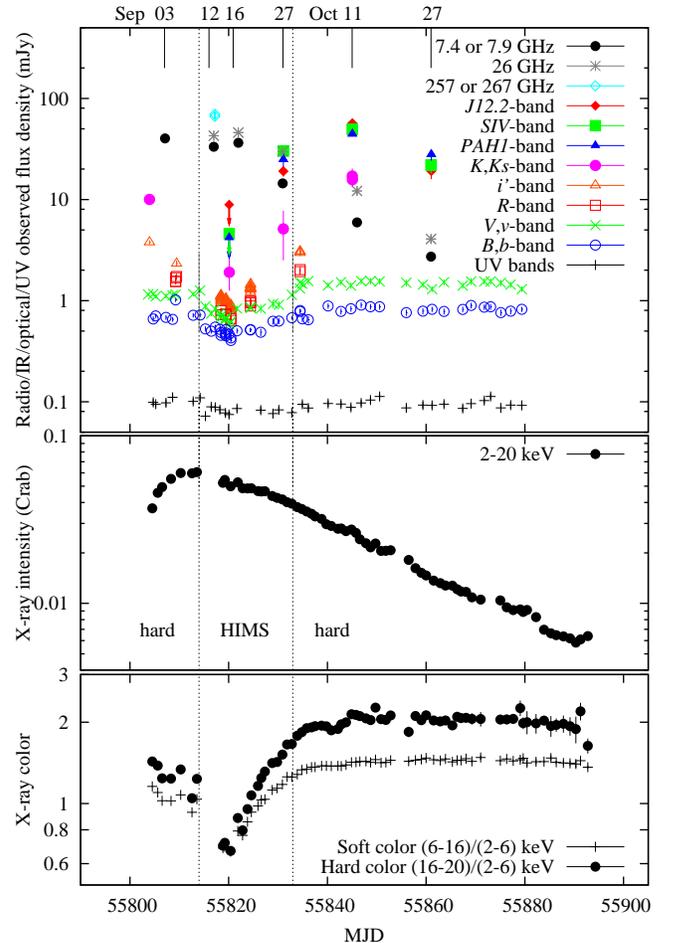}
\caption{\emph{Upper panel:} Radio, IR, optical and UV light curves of the 2011 outburst of \target. Where error bars are not visible they are smaller than the symbols. As well as VLT/VISIR mid-IR, FT optical and \textit{Swift} UVOT optical/UV data ($v$-band, $b$-band and the average flux of the three UV filters, $w1$, $m2$ and $w2$) we also include $i^{\prime}$ and $K_{\rm S}$-band fluxes on August 30 (MJD 55803) from \citeauthor{rauet11} (2011). \emph{Center panel:} X-ray (\textit{RXTE}) light curve. \emph{Lower panel:} X-ray color (two intensity ratios are given). Vertical dashed lines illustrate state transition dates \citep[from][]{ferret12}.}
\end{figure}

\begin{figure*}
\centering
\includegraphics[width=16.6cm,angle=0]{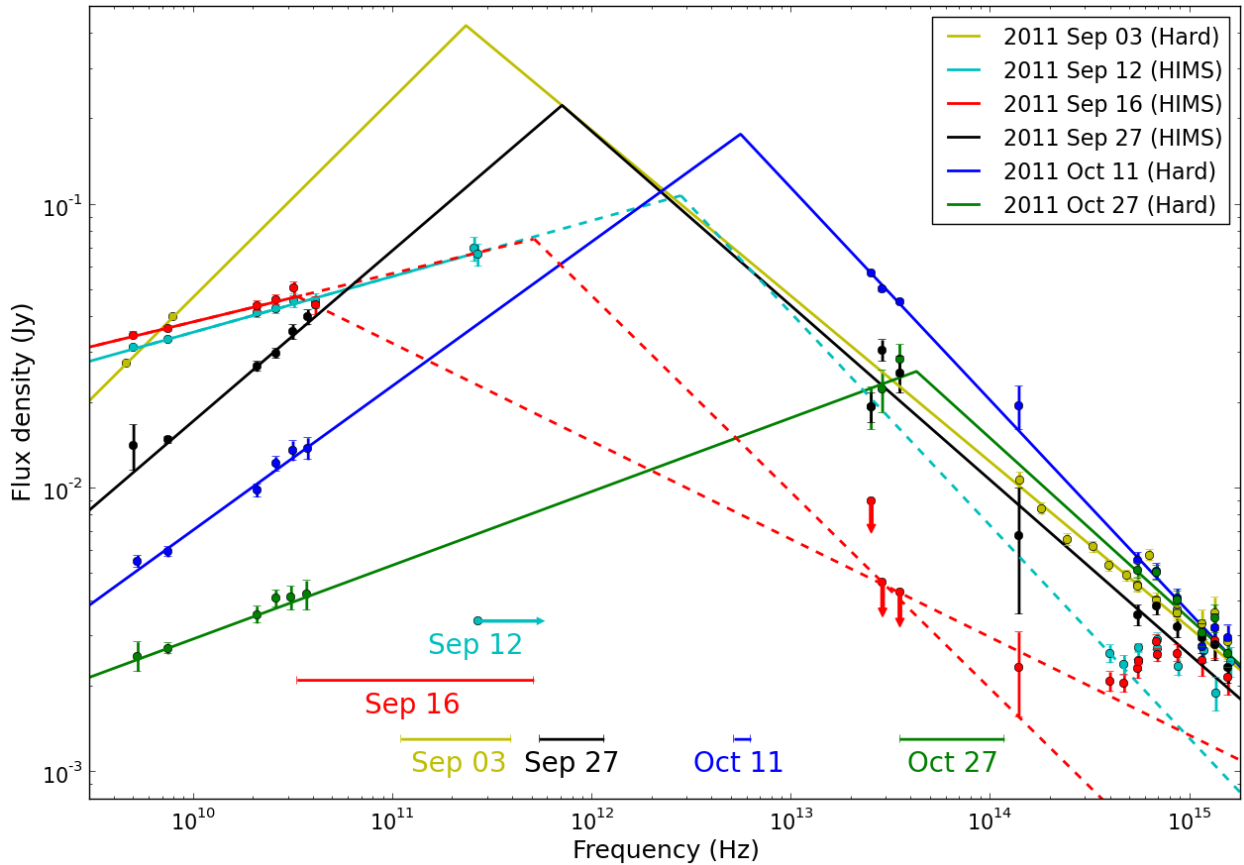}\\
\caption{Radio to UV (de-reddened) spectra with broken power law spectral fits (solid lines; see text). Dashed lines indicate possible ranges of parameters. The horizontal bars denote the uncertainty on the spectral break frequency on each date.}
\end{figure*}

\begin{figure*}
\centering
\includegraphics[width=8.0cm,angle=0]{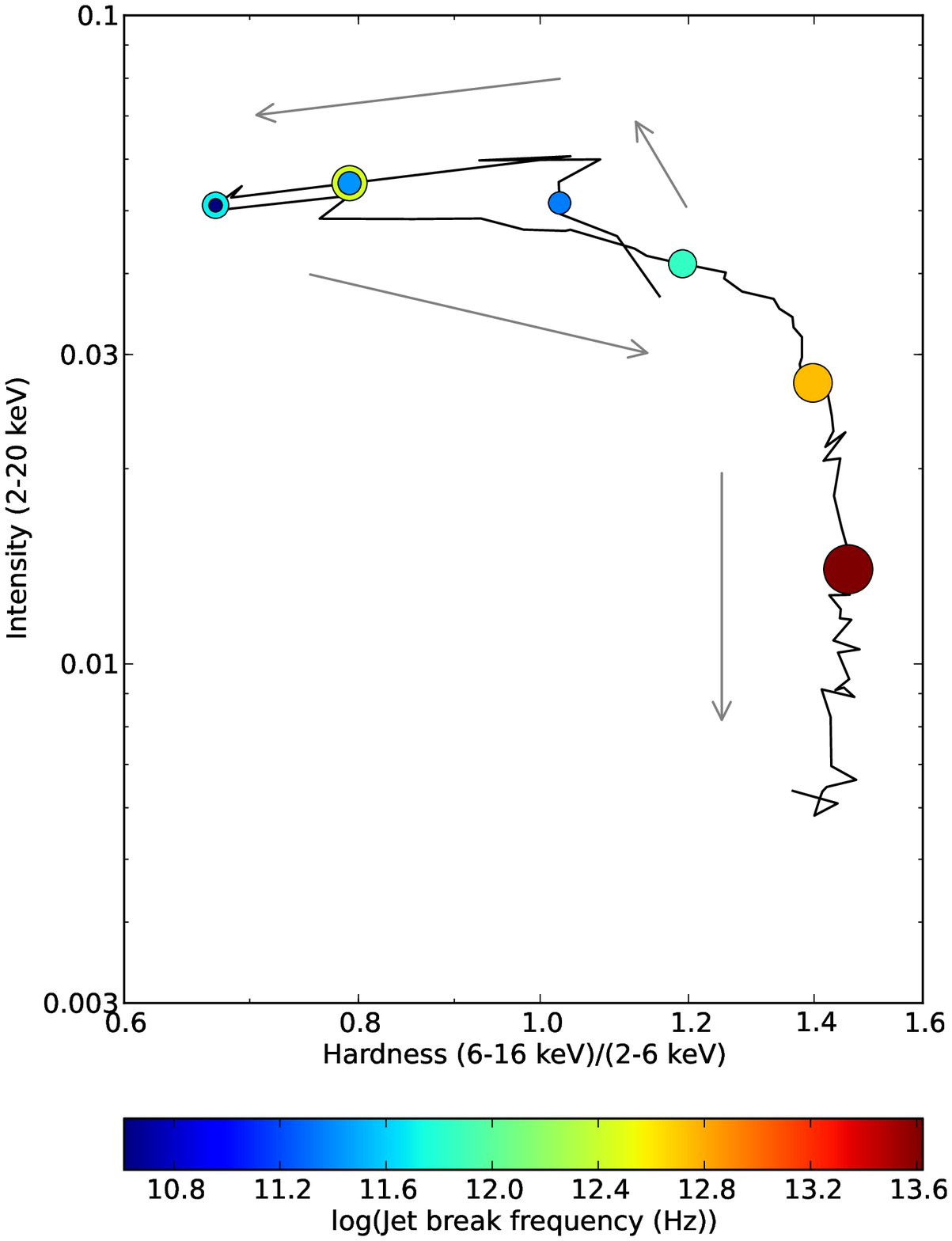}
\includegraphics[width=8.0cm,angle=0]{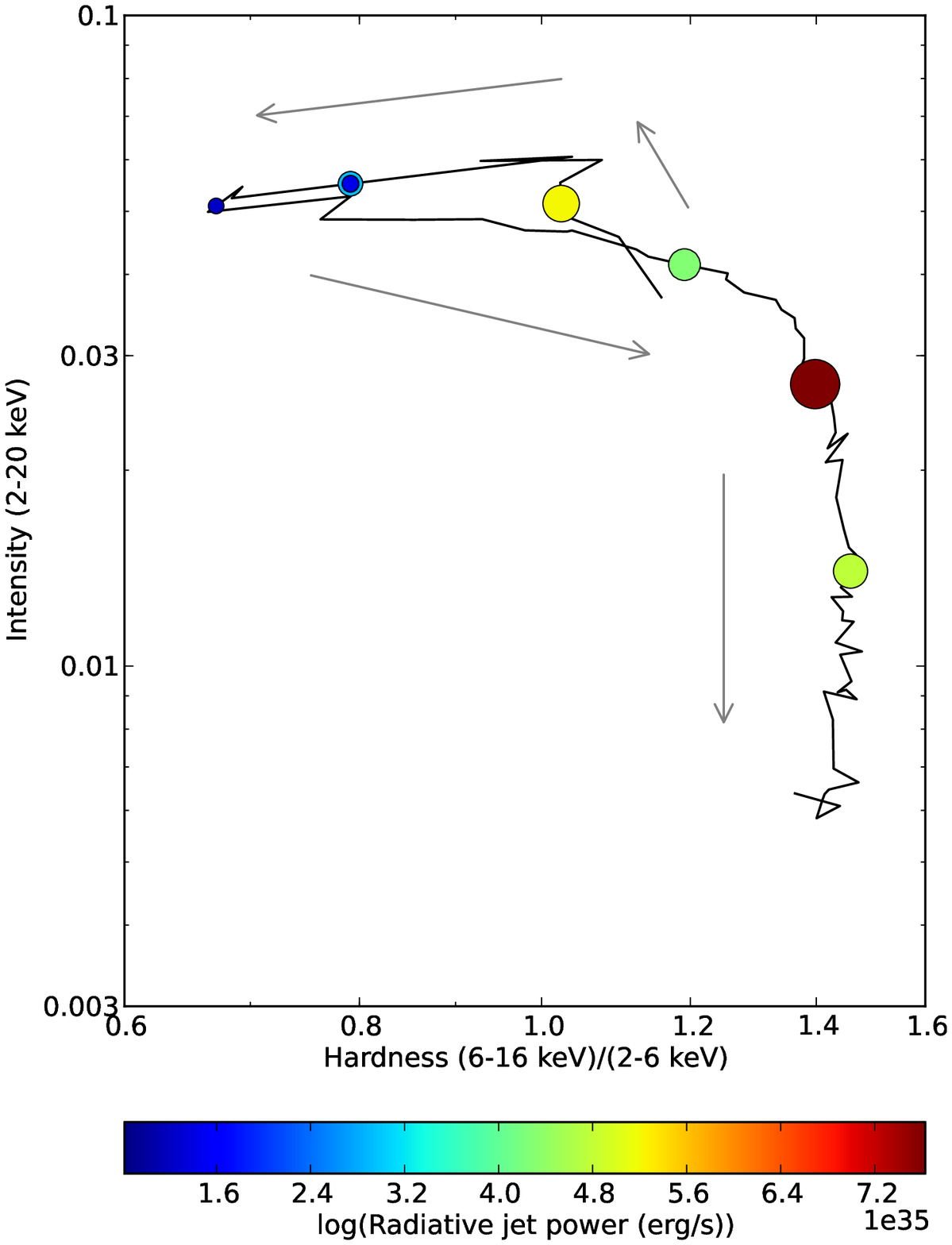}
\caption{HID of the outburst, with marker size and color indicating the inferred jet break frequency (left; double circles indicate ranges inferred) and jet luminosity (right). Arrows represent temporal evolution.}
\end{figure*}

\section{Results}

\subsection{Light curve evolution}

The multiwavelength light curves are presented in Fig. 1.
While the X-rays brightened, softened, then hardened and faded towards quiescence, the UV flux remained remarkably constant to within a factor of $\sim 2$ until November 14 \citep[In quiescence the UV \emph{w1} and \emph{w2} magnitudes are $> 3$ mag fainter than this mean outburst level;][]{yanget12b}. The optical $B$ and $V$-bands also remained approximately constant except during the intermediate states, when the source was slightly fainter. In the IR regime this dip in flux was more pronounced, and the flux was not constant after the transition. The amplitude of the flux drop/rise over the period of X-ray softening
was a factor of $> 10$ in the mid-IR filters, $\sim 10$ in $K$-band, $\sim 2.5$ in the optical $i^{\prime}$-band and $< 2$ in $B$ and the UV bands. The evolution is fairly smooth, with day-to-day rms flux variations of $\sim 10$\%.

While most BHXBs possess blue optical/IR spectra with flux densities peaking around the blue--optical/UV \citep[e.g.,][]{hyne05}, we find that the IR flux densities of \target\ are much brighter than the optical (the brightest we report is 57 mJy at 10.5 $\mu m$). Very few mid-IR detections of transient BHXBs exist in the literature, and only GRO J0422+32, GX 339--4 and Cyg X--1 (a high-mass BHXB) were detected at such bright flux densities \citep{vanpet94,fendet00,corbfe02,gandet11}.

\subsection{Broadband spectra}

In Fig. 2 we present broadband (radio to UV) spectra at six epochs in 2011 when radio and mm/IR observations were made within a few days of each other, to constrain the jet synchrotron spectrum. 
Optical/IR data were de-reddened assuming an interstellar extinction of $E(B-V) = 0.49 \pm 0.03$ ($A_{\rm V} \approx 1.5 \pm 0.1$; this is inferred using broadband radio to X-ray spectral fitting which will be presented in a follow-up publication; T. D. Russell et al., in preparation) and adopting the extinction curves of \cite*{cardet89}
and \cite{weindr01}.

The resulting broadband spectra are some of the most well-sampled in the literature for a BHXB. The radio and IR--optical spectral slopes can be approximated by power laws consistent with optically thick and optically thin synchrotron emission, respectively, confirming the dominance of the compact jet from radio to optical frequencies \citep[see also][]{millet11,rauet11,yanget12a}. The radio power law appeared more inverted than most BHXBs \citep[one similar source is XTE J1118+480;][]{fendet01}. The radio emission faded during and after the state transition, whereas the IR faded before brightening dramatically. This secondary maximum in the IR light curve is common to several BHXBs \citep[e.g.,][]{kaleet05,coriet09}.

On September 16, when the X-ray spectrum was softest, the IR jet emission had faded considerably and a separate optical component was clearly visible, which is consistent with the irradiated accretion disk (T. D. Russell et al., in preparation). However, the $K$-band detection, slightly brighter than the optical, implies the jet was still contributing, and with the mid-IR upper limits, still allows us to constrain the optically thin spectral index (shown by dashed red lines in Fig. 2).

We fitted broken power laws to the broadband spectra, to infer the flux and frequency of the spectral break (the results are given in Table 1). The fits used data up to $B$-band, on all dates except when the disk made a strong contribution to the optical emission (September 12 and 16).  The UV data were excluded from all fits because the de-reddened UV flux had much greater uncertainties due to the stronger influence of extinction and a significant but uncertain (and evolving) disk contribution. We estimate a maximum possible disk contribution of 30$\%$ in $B$-band on all dates except September 12 and 16. The systematic uncertainty arising from this possible disk contribution is estimated and propagated into the fits, resulting in larger positive errors associated with the spectral index and inferred jet break frequencies (Table 1). Other processes such as synchrotron-self Comptonization cannot bias the power law fits since their emission dominates at energies much higher than we fit here.

A dramatic evolution of the jet spectrum is evident (Fig. 2). The jet break resided at mm frequencies in the first four epochs, then shifted by more than two orders of magnitude to higher frequencies during the transition back to the hard state, being detected between our mid-IR and optical data on October 27. This shift caused an increase in the IR flux of the jet; the changing jet spectrum is responsible for the wavelength-dependent drop/rise in flux in the IR/optical light curves (Fig. 1). The observed IR spectral index, and the wavelength dependence of the drop/rise in flux over the transition, are both inconsistent with the hot accretion flow model proposed by \cite*{veleet13}.

\subsection{Relation with X-ray hardness}

The X-ray hardness--intensity diagram (HID) from \textit{RXTE} observations is presented in Fig. 3. The overlaid circles represent the frequency of the jet break (left panel) and the total radiative jet luminosity integrated over the spectrum up to $B$-band assuming a source distance of 8 kpc (right panel). The frequency and flux of the jet break provides an estimate of the minimum radiative luminosity of the jet \citep[e.g.,][]{corbfe02}. We cannot constrain the frequency of the high energy cooling break in the synchrotron spectrum, which occurs in the UV--X-ray regime, and we consider these estimates to represent lower limits on the radiative jet power. The total jet power will greatly exceed these values, as most of the power is locked in the kinetic flow of matter, and the jet radiative efficiency may be $\sim 5$ \% \citep[e.g.,][]{fend01}. Using a composite Monte Carlo Spearman's Rank correlation test, we find tentative evidence for a positive correlation between the jet break frequency and X-ray hardness (as visually apparent in Fig. 3). The Spearman's Rank coefficient is $0.79^{+0.21}_{-0.29}$, giving a significance level of $0.88^{+0.12}_{-0.22}$ for the correlation.

\section{Discussion}

We have witnessed the gradual fading and recovery of a compact jet in the BHXB \target, when it underwent transitions between the hard state and the HIMS. This suggests that the stability of the jet changes when the source moves away from the hard state, likely becoming less powerful.
The results suggest the jet power is sensitive to the changing structure of the accretion flow between the canonical hard state and the HIMS. In the hard state, theoretically predicted scaling relations between inflow and outflow are observed empirically, at least in some BHXBs \citep[e.g.,][and references therein]{gallet12,corbet13a,russet13}. These relations assume that a constant fraction of the matter being accreted is channeled into the jet, with total jet power, $P_{\rm jet} \propto \dot{m}_{\rm jet} \propto \dot{m}_{\rm disk}$ (where $\dot{m}$ is the mass accretion rate). In the HIMS the radio luminosity of \target\ was not an accurate tracer of the total radiative jet luminosity because the latter decreased while the radio remained bright. The hard state scaling relations therefore no longer hold in the HIMS. The radiative jet luminosity (up to $7 \times 10^{14}$ Hz) increased modestly with X-ray hardness (Fig. 3). The ratio $\dot{m}_{\rm jet} / \dot{m}_{\rm disk}$ was therefore reduced in this state compared to the hard state. The jet power probably made up an increasing fraction of the total energy budget as the spectrum hardened and the outburst decayed (unseen changes in the frequency of the high energy synchrotron cooling break could affect this). The jet appeared less powerful in the HIMS but did not quench or recover immediately. If the jet switched off/on instantaneously we would have observed a decrease/increase in its luminosity on short timescales -- at least as long as the light crossing time over the photosphere or emitting region, and at most the adiabatic expansion timescale or the synchrotron cooling timescale of the propagating matter in the jet \citep*[the latter depends on the magnetic field strength, and is on the order of minutes for IR and days for radio frequencies, adopting $B = 10^{4}$ G;][]{millet04,gandet11}. Instead, the jet power varied on timescales of weeks, the same as the evolving inner accretion flow.

\small
\begin{table*}
\begin{center}
\caption{Results of the power law spectral fitting, and inferred jet breaks and jet luminosities. For spectral indices, the convention $F_{\nu} \propto \nu^{\alpha}$ is adopted. The error bars are 90\% confidence limits. The integrated ($5 \times 10^9$ Hz to $7 \times 10^{14}$ Hz) jet luminosities assume a source distance of 8 kpc.}
\vspace{-1.5mm}
\begin{tabular}{ccccccc}
\hline
Date & MJD (UT) & Optically thick PL, & Optically thin PL, & Break frequency, & $\rm Flux_{break}$ & $\rm L_{jet}$ \\ 
 (VLA) & (VLA) & $\alpha_{\rm thick}$ & $\alpha_{\rm thin}$ & $\nu_{\rm b}$ (Hz) &  (mJy) & (erg s$^{-1}$) \\
\hline
2011 Sep 03 & 55807.12 & $0.70 \pm 0.10$ & $-0.60 \pm 0.10$ & $2.30_{-1.20}^{+1.62}\times10^{11}$ & $426_{-178}^{+697}$ & $5.21^{+2.23}_{-1.44} \times 10^{35}$ \\ 
2011 Sep 12 & 55816.97 & $0.20 \pm 0.01$ & $\leq -0.45$ & $\geq 2.67 \times 10^{11}$ & $\geq 68$ & (1.45 -- 2.94) $\times 10^{35}$ \\
2011 Sep 17 & 55821.97 & $0.17 \pm 0.02$ & $-0.69$ -- $-0.35$ &(0.410 -- 5.08) $\times10^{11}$ & 47 -- 76 & (0.82 -- 1.23) $\times 10^{35}$ \\
2011 Sep 26 & 55830.95 & $0.60_{-0.03}^{+0.02}$ & $-0.62_{-0.03}^{+0.08}$ & $7.04_{-1.60}^{+4.46}\times10^{11}$ & $221_{-39}^{+121}$ & $4.30^{+1.31}_{-0.42} \times 10^{35}$ \\
2011 Oct 12 & 55846.01 & $0.51 \pm 0.01$ & $-0.75_{-0.05}^{+0.06}$ & $5.54_{-0.38}^{+0.73}\times10^{12}$ & $180_{-15}^{+29}$ & $7.63^{+1.54}_{-1.09} \times 10^{35}$ \\
2011 Oct 27 & 55861.00 & $0.26 \pm 0.02$ & $-0.64_{-0.04}^{+0.12}$ & $4.20_{-0.71}^{+7.53}\times10^{13}$ & $26_{-5}^{+17}$ & $4.77^{+1.50}_{-0.49} \times 10^{35}$ \\
\hline
\end{tabular}
\end{center}
\end{table*}
\normalsize

Compact jets are extinguished (followed by a bright, discrete ejection) either around the transition between the HIMS and the soft--intermediate state (the `jet line' in the HID) or before this, associated with a drop in the X-ray variability, and they tend to reactivate as the source hardens again in the HIMS \citep[e.g.,][]{fendet09,millet12,corbet13b}. The jet in \target\ was never extinguished completely, and the source never entered the soft--intermediate state, which is normally associated with the quenching of the compact jet and the launching of discrete ejections.
The jet fading/recovering occurred in the HIMS, so the process was triggered by some activity of this state. Generally in the HIMS, the X-ray power law is softer, the disk more prominent and the fractional rms variability is decreased, compared to the hard state \citep{bell10}. The disk inner radius is smaller in the HIMS \citep{ferret12}, perhaps preventing as much matter from entering the jet nozzle, decreasing the $\dot{m}_{\rm jet} / \dot{m}_{\rm disk}$ ratio, in comparison with the hard state. Thermal disc winds are detected in the HIMS and most strongly in the soft state \citep{neille09,pontet12}, so our results are consistent with the disc wind and jet accretion powers being inversely related; the wind may help quench the jet \citep[see also][]{milleret12}. Whatever the physical process is, these observations can provide vital constraints for models of the jet production mechanism.

A further, interesting constraint can be made from how the jet luminosity increase manifests itself over the HIMS to hard state transition. While the IR brightened considerably, the radio faded slightly and the jet break shifted to higher frequencies.
These phenomena could be explained if the size scale of the jet base, the magnetic field strength or the particle pressure increased \citep{heinsu03}. In this scenario the IR flux and jet break frequency would increase while the radio emission would not be affected significantly, if the jet opening angle is constant.
In addition, the gradual increase in the jet luminosity as the source hardened is consistent with the delay in the IR brightening after the transition is completed, seen in a number of BHXBs \citep[e.g.,][]{kaleet05,coriet09,russet10,corbet13b}. It is not clear why the radio spectrum evolved from one which was slightly inverted when the X-rays were softest ($\alpha \sim 0.2$, where $F_{\nu} \propto \nu^{\alpha}$), to a very inverted spectrum as it hardened again ($\alpha \sim 0.5$). However, these results are consistent with the spectrum of compact jets becoming optically thin when they switch off \citep*{fendet04,millet12} and flattening when they switch back on \citep{corbet13a,corbet13b}.

For the first time, the detailed changes of the energetics of a relativistic jet have been witnessed over a state transition. In time these observations can place stringent constraints on time-dependent jet models, providing a step forward in our understanding of how jets are launched. These results also demonstrate that ground-based mid-IR monitoring of XBs can open up a new window into jet physics in these systems, especially when coordinated with other wavelengths. Very few sub-mm observations of BHXBs exist, but the measured $\sim 70$ mJy sub-mm fluxes of \target\ and implied fluxes exceeding 0.1 Jy (Fig. 2) suggest future outbursts of BHXBs will be easily detected by several current sub-mm facilities.

\acknowledgments

DMR would like to thank Mario van den Ancker and Christian Hummel at ESO for help with the preparation and execution of the VLT/VISIR observations. DMR acknowledges support from a Marie Curie Intra European Fellowship within the 7th European Community Framework Programme (FP7) under contract no. IEF 274805. This work was supported by Australian Research Council grant DP120102393, FP7 grant agreement number ITN 215212 Black Hole Universe, the Spanish Ministry of Economy and Competitiveness (MINECO) project AYA2010-18080, and ANR ``CHAOS'' (ANR-12-BS05-0009). The National Radio Astronomy Observatory is a facility of the National Science Foundation operated under cooperative agreement by Associated Universities, Inc. The Submillimeter Array is a joint project between the Smithsonian Astrophysical Observatory and the Academia Sinica Institute of Astronomy and Astrophysics and is funded by the Smithsonian Institution and the Academia Sinica. The Faulkes Telescope South is maintained and operated by Las Cumbres Observatory Global Telescope Network.

\end{document}